# Giant Surface-driven Nonlinear Hall Effect in BiTeCl at Room Temperature

Zhihua Liu[1,2], Ziheng Wang[3], Yongbo Lv[4], Hanru Feng[1], Zhiwei Zhang[1], Bo Zhang[1], Feng Liu[1], Guohua Wang[1], Shengwei Jiang[1], Hao Chu[4], Hui Li[3,*], Dong Qian[1,2,5,*]

[1]State Key Laboratory of Micro-nano Engineering Science, Key Laboratory of Artificial Structures and Quantum Control (Ministry of Education), School of Physics and Astronomy, Shanghai Jiao Tong University, Shanghai 200240, China

[2]Tsung-Dao Lee Institute, Shanghai Jiao Tong University, Shanghai 200240, China

[3]College of Engineering Physics, Shenzhen Technology University, Shenzhen 518118, China

[4]Center for Ultrafast Science and Technology, School of Physics and Astronomy, Shanghai Jiao Tong University, Shanghai 200240, China

[5]Collaborative Innovation Center of Advanced Microstructures, Nanjing 210093, China

## Abstract

The nonlinear Hall effect (NLHE) provides a pathway to generate a Hall response in time-reversal-symmetric yet inversion-symmetry-broken systems. NLHE can rectify an alternating current into a transverse direct voltage, making it attractive for radio-frequency rectification, energy harvesting, and terahertz detection, applications for which device miniaturization remains a central pursuit. In this context, the inherent inversion symmetry breaking at surfaces is particularly appealing: because symmetry is necessarily broken at the surface of any crystal, irrespective of whether its bulk is centrosymmetric, surface-driven nonlinear responses lift the stringent constraint on bulk symmetry and open a route toward compact device architectures. Here we report the observation of a giant, surface-driven second-order nonlinear Hall effect in the Rashba-type polar semiconductor BiTeCl at room temperature. The determined second-order nonlinear Hall susceptibility at 300 K reaches 1.68 $\mu mV^{-1}$, which is 80 times larger than that of the best previously reported surface-dominated systems. We attribute this giant response to the synergistic interplay between BiTeCl's polar crystal structure and its rich surface states: the polar stacking renders the top and bottom surfaces inequivalent, so that the nonlinear response originates from a single surface without compensation from the other. Symmetry and scaling analyses suggest that both skew-scattering and side-jump mechanisms contribute to the observed effect. Our findings not only identify BiTeCl as a promising platform for future applications utilizing the NLHE, but also establish the asymmetry between the opposite surfaces of a polar crystal as a general design principle for discovering surface-driven materials with larger nonlinear Hall responses.

*E-mail: lihui@sztu.edu.cn (Hui Li); dqian@sjtu.edu.cn (Dong Qian).

## 1 Introduction

The Hall effect typically occurs in systems with broken time-reversal symmetry. In the ordinary Hall effect, this symmetry breaking is induced by an external magnetic field, whereas in the anomalous Hall effect, it arises from the spontaneous magnetic order. Recently, studies on the second-order nonlinear Hall effect (NLHE) have demonstrated that an intrinsic nonlinear Hall response can arise in non-magnetic materials with broken inversion symmetry in the absence of an external magnetic field, and the effect is proportional to the Berry curvature dipole (BCD) [1, 2]. In magnetic systems with broken time-reversal symmetry, a distinct intrinsic contribution can arise from the quantum metric dipole [3–6]. This effect provides potential applications in energy harvest [7, 8], wireless communications and wide-band rectification [9–11]. However, the symmetry constraints associated with a non-vanishing BCD limit its observation to a limited number of crystal point groups of crystal point groups [12]. In contrast, extrinsic scattering processes, such as skew scattering and side-jump effects [12–16], provide alternative mechanisms for the second-order NLHE that can occur in crystals with broken inversion symmetry [12–17]. More generally, although not all crystals break inversion symmetry in their bulk, inversion symmetry is inherently broken at surfaces or interfaces. Therefore, surface/interface-driven nonlinear Hall responses may occur in centrosymmetric crystals [18].

For example, centrosymmetric materials such as Bi [18], $Bi_2Se_3$ [19, 20], $Bi_2Te_3$ [21, 22], $Pb1_{-x}Sn_xTe$ [23] and $LaAlO_3/KTaO_3(111)$ [24] exhibit surface/interface state dominated second-order NLHE, where the nonlinear Hall response arises from the skew scattering of electrons in surface/interface states. In $T_d$-$TaIrTe_4$ and $NbIrTe_4$, the observed second-order NLHE was proposed to arise from the nonzero BCD of the bulk states near the surface [7, 11]. In Te, the second-order NLHE originates from skew scattering of bulk electrons near the surface [25]. Artificial stacking provides another effective route for breaking inversion symmetry in van der Waals systems. A representative example is the hBN/graphene/hBN heterostructure, where the moiré potential between graphene and hBN breaks inversion symmetry [15]. Following this strategy, giant nonlinear Hall responses have been reported in twisted bilayer graphene [26, 27], twisted double bilayer graphene [28, 29], twisted $WSe_2$ [30], and $WTe_2/WSe_2$ heterostructures [31]. These results demonstrate the power of stacking and twisting in engineering symmetry-broken electronic states. Nevertheless, such artificial structures generally rely on delicate transfer processes, precise twist-angle or stacking control, and often require low-temperature operation, which poses substantial challenges for scalable applications. Therefore, discovering naturally available material platforms with large room-temperature nonlinear Hall responses and simple device fabrication remains an important goal. In this context, surface-driven nonlinear Hall materials offer a promising direction.

However, the reported surface-driven second-order nonlinear Hall susceptibility ($\chi_{xy}^{2\omega}$) is significantly smaller than that achieved in bulk non-centrosymmetric systems [8]. Therefore, it is interesting and highly desirable to explore materials exhibiting surface-dominated NLHE with large nonlinear Hall susceptibility.

The relatively small nonlinear Hall susceptibility reported in previous surface-driven systems may be partly attributed to the cancellation between the top and bottom surface contributions. Although both surfaces can generate second-order nonlinear Hall responses due to local inversion-symmetry breaking, their structural polarities are generally opposite. As a result, the corresponding nonlinear Hall contributions from the two surfaces may partially cancel each other in a global transport measurement, leading to a weak net surface nonlinear Hall signal. In polar crystals, such cancellation can be reduced or avoided because the two surfaces are intrinsically inequivalent. An interesting example is BiTeBr, a layered polar crystal with a non-centrosymmetric structure (space group P3m1, No. 156), which hosts surface states with large Rashba splitting [32]. A large second-order nonlinear Hall response persisting above room temperature has been reported in BiTeBr, involving both surface and bulk contributions [33]. However, a quantitative determination of the surface contribution has not yet been achieved. Analogous to BiTeBr, BiTeCl (space group $P6_3mc$, No. 186) is also a layered polar crystal and hosts multiple Rashba-split surface states [34–38]. Importantly, owing to its six-fold bulk crystal symmetry, the in-plane bulk second-order nonlinear Hall response is symmetry-forbidden in BiTeCl [12]. These features make BiTeCl a promising candidate for realizing a large nonlinear Hall response dominated by surface states.

In this work, we report the observation of a large, in-plane, surface-driven second-order NLHE in BiTeCl. The measured $\chi_{xy}^{2\omega}$ is 80 times larger than that of the best previously reported surface-dominated systems. Our results establish BiTeCl as a promising platform for achieving strong nonlinear electrical responses, with a surface-dominated origin, which is particularly advantageous for device miniaturization.

## 2 Materials and methods

BiTeCl single crystals were grown via a two-step self-flux method. In the first step, $Bi_2Te_3$ precursors were synthesized by heating a stoichiometric mixture of Bi (99.997%) and Te (99.999%) powders to 950 °C, followed by slow cooling to 550 °C with intermediate annealing. Subsequently, the as-prepared $Bi_2Te_3$ crystals were mixed with $BiCl_3$ flux at a molar ratio of $Bi_2Te_3$:$BiCl_3$ = 1:8, heated to 440 °C, then slowly cooled to 200 °C over 90 h, followed by a final annealing step for 24 h [39]. This protocol successfully produced BiTeCl single crystals with lateral dimensions of several millimeters. The non-centrosymmetric crystal structure of BiTeCl [40] is shown in Fig. 1(a). One unit cell includes two Bi-Te-Cl trilayers (TLs). Along the *c*-axis, BiTeCl crystal has $6mm$ symmetry, whereas a single Bi-Te-Cl TL exhibits $3m$ symmetry. Natural cleavage in BiTeCl occurs within the *ab*-plane. The structure of the grown crystals was checked by X-ray diffraction (XRD) and Laue backscattering. As shown in Fig. 1(b), the single-crystal XRD spectrum of the grown crystals exhibits a series of sharp (0,0,2$l$) Bragg peaks, yielding a lattice constant c = 1.235 nm, consistent with previous report [40]. Furthermore, the sharp Laue backscattering pattern (Fig. 1(c)) also agrees well with the simulation results (Fig. 1(d)).

BiTeCl flakes were mechanically exfoliated onto $SiO_2$(285 nm)/Si substrates. Pre-patterned Au electrodes were transferred onto selected flakes, followed by spin coating with photoresist (AZ 1500). These steps were conducted in an argon-filled glove box to prevent degradation. The devices were then patterned via laser direct writing lithography (MicroWriter ML3, Quantum Design). The exposed regions were etched by reactive ion etching (RIE) using argon at a base pressure of $2.0\times10^{-4}$ Torr. Finally, the photoresist was removed by soaking in N-methyl-2-pyrrolidone (NMP), and a protective layer of polymethyl methacrylate (PMMA) was spin-coated onto the device. Electrical transport measurements were carried out using a physical property measurement system (PPMS, Quantum Design). External lock-in amplifiers (Stanford Research Systems SR830 and Guangzhou Sine Scientific Instrument OE1022) were used.

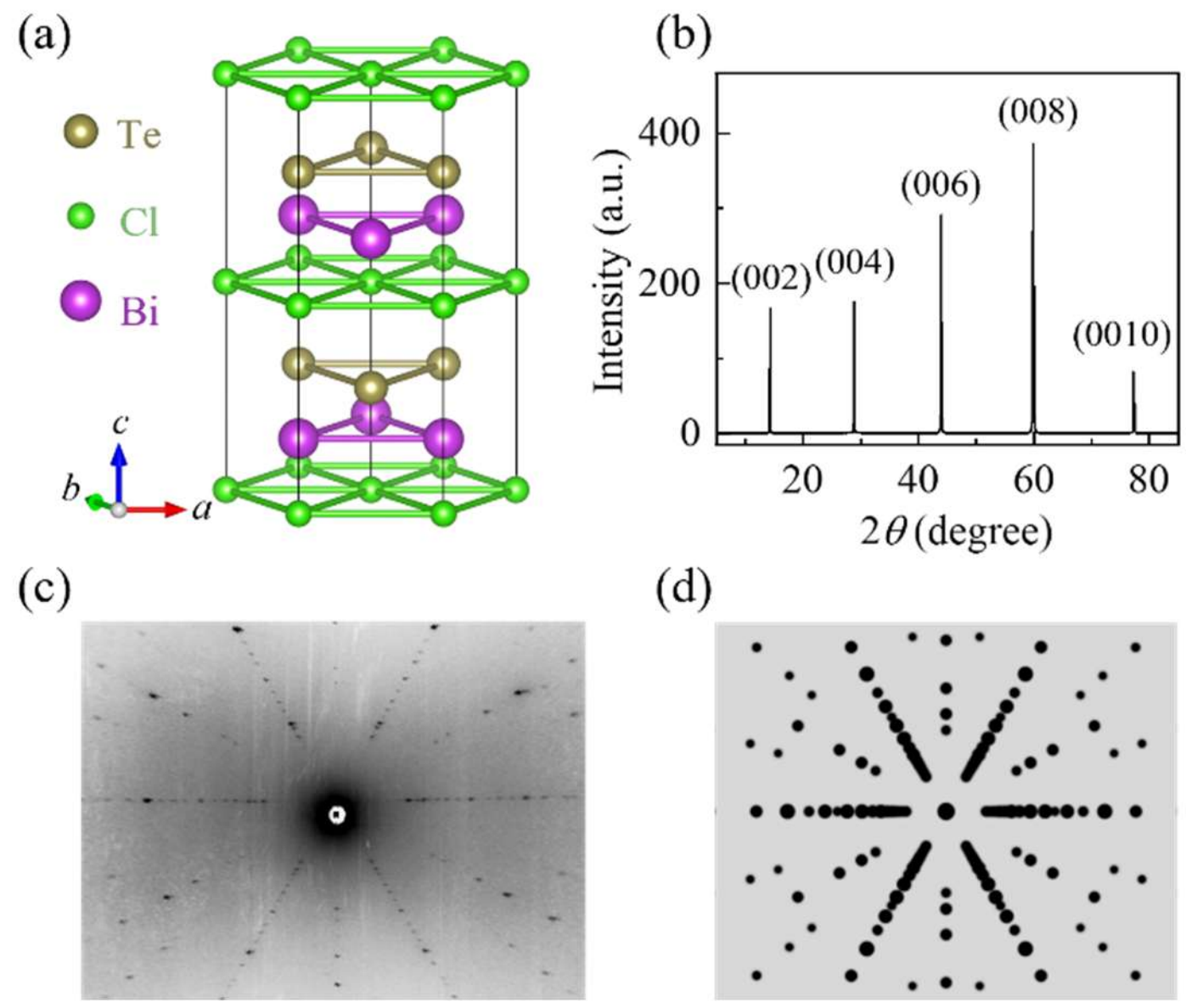

**Figure 1** Structural Characterization of BiTeCl. (a) Schematic of the crystal structure. (b) Single-crystal XRD spectrum shows sharp (0,0,2*l*) diffraction peaks. (c) X-ray Laue backscattering pattern with the X-ray incident along the [001] direction. (d) Simulated Laue backscattering pattern.

**3 Results and discussion**

As shown in the inset of Fig. 2(a), the selected BiTeCl flakes were patterned into a circular shape, and 12 Au electrodes were fabricated to suppress geometry-induced artifacts. Although stoichiometric BiTeCl should be a narrow-gap semiconductor [36, 37], the temperature-dependent resistivity of a 12-nm-thick BiTeCl flake (Fig. 2(a)) exhibits metallic behavior due to Te and Cl vacancies and Bi/Te antisite defects [38]. Hall resistivity ($\rho_{xy}$) as a function of the magnetic field applied along the out-of-plane direction is shown in Fig. 2(b), yielding a carrier density of $n = 5.88\times10^{18}$ cm$^{-3}$, and carrier mobility $\mu = 669$ cm$^2$V$^{-1}$s$^{-1}$, consistent with previous reports [38].

Furthermore, we systematically investigated the nonlinear electrical transport properties of this 12 nm BiTeCl flake in the absence of an external magnetic field. We applied an AC current $I(t) = I_0 sin\omega t$ to the sample and simultaneously measured the first- and second-order longitudinal and transverse voltages ($V_{xx}^{\omega}$, $V_{xy}^{\omega}$, $V_{xx}^{2\omega}$, $V_{xy}^{2\omega}$) at room temperature. According to the crystal symmetry of BiTeCl, $V_{xy}^{\omega}$ should be zero. Any finite $V_{xy}^{\omega}$ observed is attributed to imperfect electrode alignment. Theoretically, the transverse voltage $V_{xy}(t)$ obeys the relation [2, 20, 41]: $V_{xy}(t) \propto \chi_{xy}^{2\omega}[I_0^2 + I_0^2\sin(2\omega t - \pi/2)]$, where $V_{xy}^{2\omega}(t) \propto I_0^2\sin(2\omega t - \pi/2)$. Therefore, two criteria can be used to validate the reliability of the measurements. First, $V_{xy}^{2\omega}$ should be proportional to the square of the amplitude of the applied current. Second, $V_{xy}^{2\omega}$ should change sign when the current direction and the corresponding Hall probe are reversed. As shown in Fig. 2(c), $V_{xy}^{2\omega}$ indeed exhibits a linearly dependence $I^2$. It should be noted, however, that thermal effects—such as the thermoelectric or Nernst response driven by Joule heating—also scale quadratically with the applied current and could in principle mimic a second-harmonic transverse signal. We therefore examined this possibility in detail and find that a thermal origin cannot account for the magnitude and behavior of the observed $V_{xy}^{2\omega}$ (see Supplementary material Note I for details). In Fig. 2(c), two measurement geometries were used, labeled by blue and red colors. The applied current and the corresponding Hall probe are reversed between the two geometries. Moreover, the detected $V_{xy}^{2\omega}$ changes sign upon reversing the direction of the applied current and Hall probe. In addition, $V_{xy}^{2\omega}$ is nearly independent of $f$ (see Supplementary material Note II for details), which rules out artifacts such as capacitive coupling [42]. These observations unambiguously demonstrate the presence of in-plane second-harmonic NLHE in BiTeCl at room temperature.

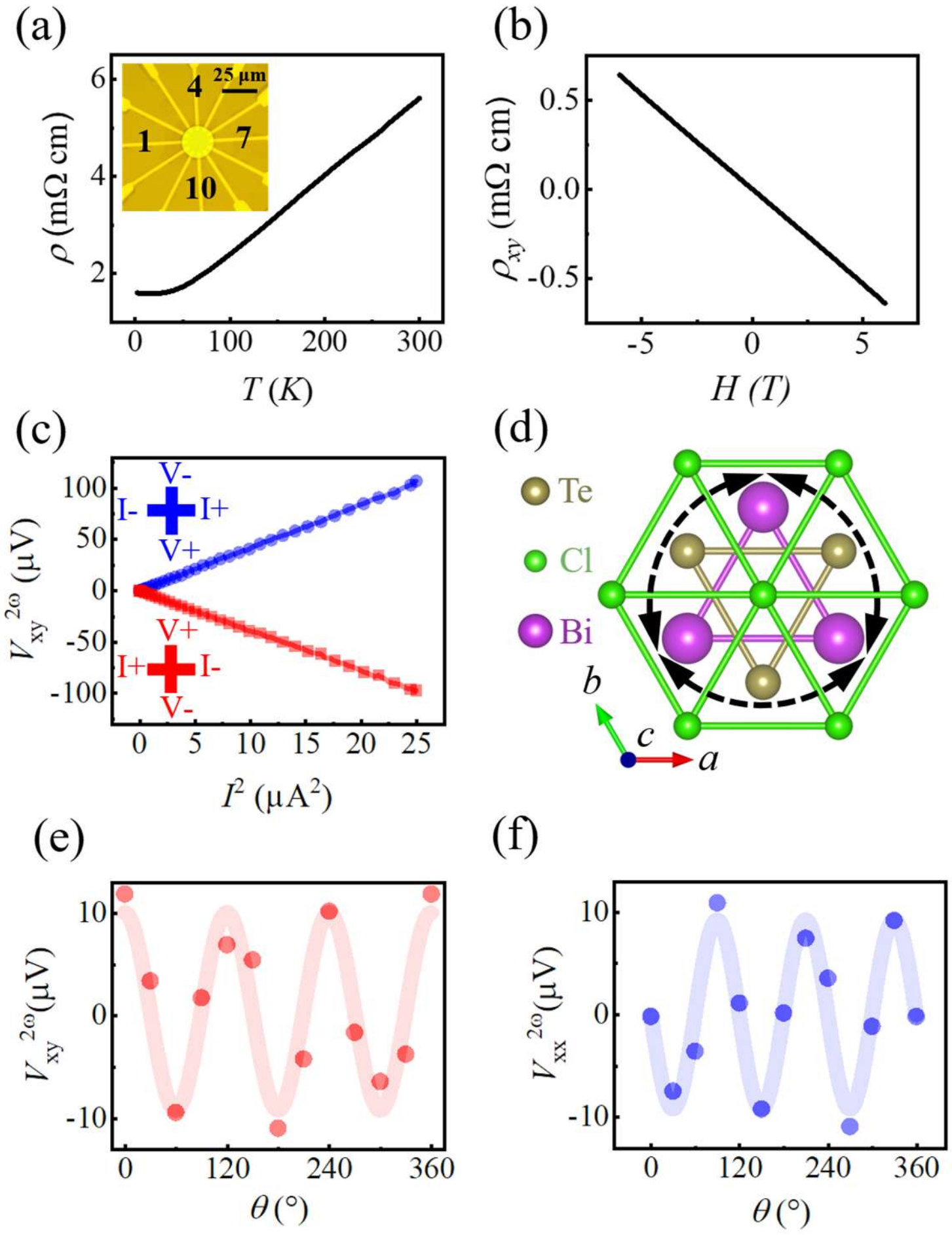

**Figure 2** Transport properties of a 12-nm-thick BiTeCl flake. (a) Resistivity as a function of temperature. Inset is the optical image of the device. Current is applied between electrodes 1 and 7. (b) $\rho_{xy}$ as a function of magnetic field applied along the $c$-direction. (c) $V_{xy}^{2\omega}$ as a function of $I^2$ measured in two measurement configurations at 300 K. (d) Top view of a Bi-Te-Cl TL. (e) $V_{xy}^{2\omega}$ and (f) $V_{xx}^{2\omega}$ as a function of in-plane azimuthal angle $\theta$ measured at 300 K. $\theta$ = 0 corresponds to the [010] axis, applied current $I$ = 2.2 μA. $f$ = 17.777 Hz.

The crystal symmetry of BiTeCl forbids a bulk second-order nonlinear Hall response. The bulk crystal belongs to the 6*mm* point group, in which all in-plane components of the second-order conductivity tensor vanish [12]. However, nonzero $V_{xy}^{2\omega}$ can originate from the surface contribution. Considering a surface Bi-Te-Cl TL, the symmetry is reduced to 3*m* instead of 6*mm*, as illustrated in Fig. 2(d). This lower symmetry allows nonzero extrinsic in-plane second-order Hall responses, characterized by a threefold in-plane azimuthal angular dependence (see Supplementary material Note III for details) [42]. To confirm this scenario, we performed in-plane angle-dependent measurements of $V_{xy}^{2\omega}$ and $V_{xx}^{2\omega}$, as shown in Figs. 2(e) and 2(f) (see Supplementary material Note IV for details). Indeed, both $V_{xy}^{2\omega}$ and $V_{xx}^{2\omega}$ exhibit a threefold rotational symmetry. The phase relationship between $V_{xy}^{2\omega}$ and $V_{xx}^{2\omega}$ is also consistent with the nonlinear conductivity tensor constraint $\sigma_{yxx}^{(2)} = -\sigma_{yyy}^{(2)}$, as required by the 3*m* point-group symmetry [12]. Therefore, we can conclude that the observed second-order NLHE is generated by the surface Bi-Te-Cl TL.

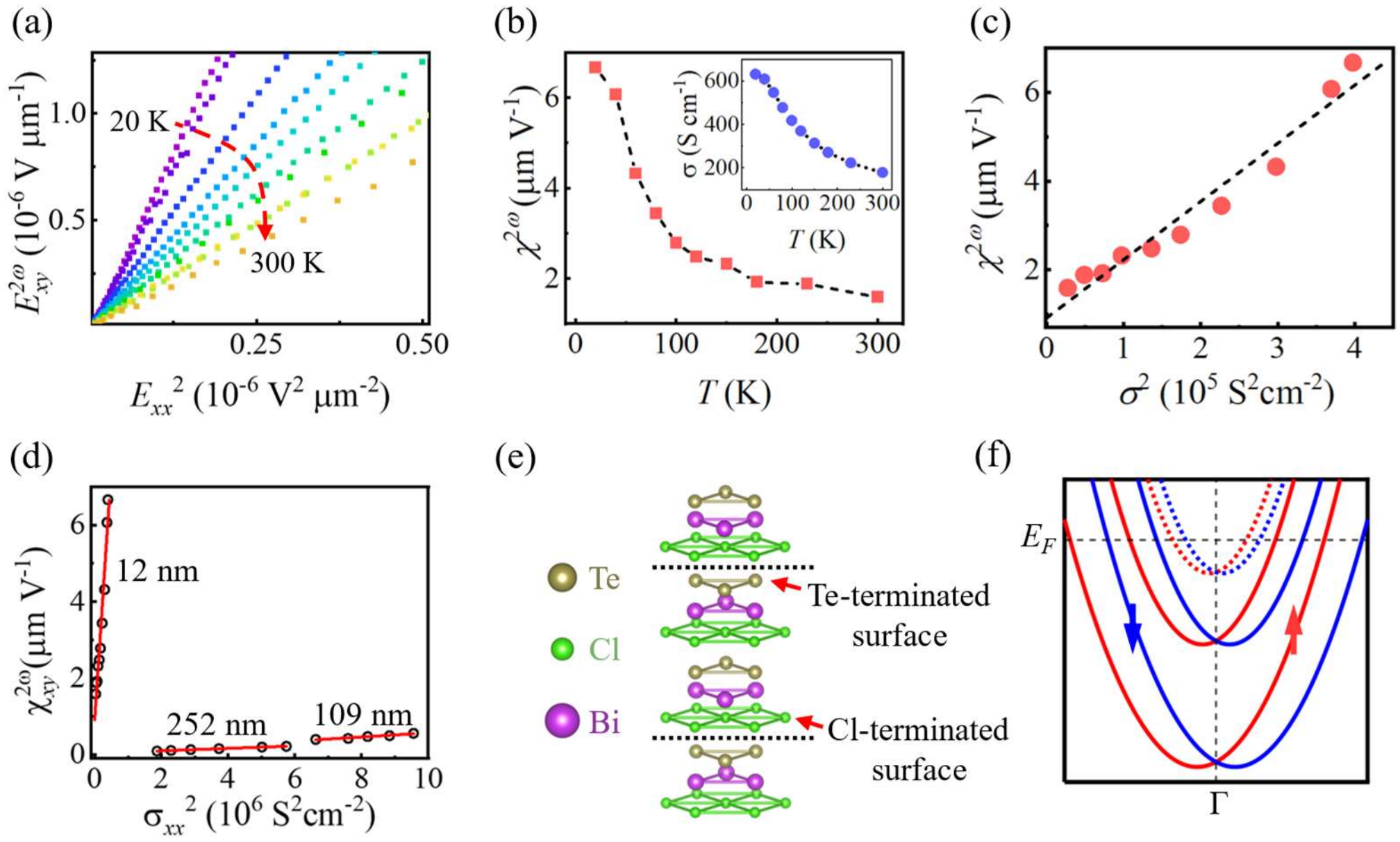

**Figure 3** Mechanism of giant surface-driven NLHE in BiTeCl. (a) $E_{xy}^{2\omega}$ as a function of $E_{xx}^2$ from 20 K to 300 K. (b) Temperature evolution of $\chi_{xy}^{2\omega}$. Inset is the temperature-dependent conductivity $\sigma_{xx}$. (c) $\chi_{xy}^{2\omega}$ as a function of $\sigma_{xx}^2$. (d) $\chi_{xy}^{2\omega}$ as a function of $\sigma_{xx}^2$ for different sample thicknesses. (e) Schematic illustration of the asymmetric surfaces after cleavage (dashed lines indicating the possible cleaving planes). (f) Schematic illustration of the electronic bands of BiTeCl near the Fermi level on the Te-terminated surface. Solid lines are two Rashba-split surface state bands. Dashed lines are surface resonance bands. Red and blue colors represent opposite spin orientations.

We define $E_{xx} \equiv V_{xx}/L_x$ and $E_{xy}^{2\omega} \equiv V_{xy}^{2\omega}/L_y$, where $L_x$ and $L_y$ denote the longitudinal and transverse lengths, respectively. In our devices, $L_x = L_y = 14$ μm. As shown in Fig. 3(a), $E_{xy}^{2\omega}$ depends linearly on $E_{xx}^2$ across temperatures ranging from 20 K to 300 K. Consequently, we extract $\chi_{xy}^{2\omega}$ ($= \Delta E_{xy}^{2\omega}/\Delta {E_{xx}}^2$) by performing linear fittings to the curves in Fig. 3(a). As depicted in Fig. 3(b), the experimentally determined $\chi_{xy}^{2\omega}$ decreases with increasing temperature. Theoretically, $\chi_{xy}^{2\omega}$ follows the relation $\chi_{xy}^{2\omega} = \xi {\sigma_{xx}}^2 + \eta$, where $\xi$ quantifies the skew-scattering contribution, while $\eta$ quantifies the side-jump and BCD contributions [1, 12, 13]. For BiTeCl, the BCD contribution is absent due to symmetry constraints, thus $\eta$ corresponds solely to the side-jump mechanism. The temperature dependence of $\sigma_{xx}$ is presented in the inset of Fig. 3(b). In Fig. 3(c), we plot $\chi_{xy}^{2\omega}$ as a function of ${\sigma_{xx}}^2$. Indeed, $\chi_{xy}^{2\omega}$ exhibits a linear scaling with ${\sigma_{xx}}^2$, where the dashed line is the linear fit. We extracted $\xi = (1.31 \pm 0.10) \times 10^3 \mu m^3\ V^{-1}\ S^{-2}$ and $\eta = (0.91 \pm 0.22)\ \mu m\ V^{-1}$. At 300 K, $\chi_{xy}^{2\omega} = 1.58\ \pm 0.01\ \mu m\ V^{-1}$, indicating that the side-jump contribution is slightly larger than the skew-scattering contribution. The skew-scattering contribution increases as the temperature decreases. At 20 K, $\chi_{xy}^{2\omega} = 6.67\ \mu m\ V^{-1}$, yielding a ratio between the skew-scattering and side-jump contributions of about 5:1. Furthermore, we investigated the thickness dependence of the scaling relation. As presented in Fig. 3(d), while the linear relationship between $\chi_{xy}^{2\omega}$ and ${\sigma_{xx}}^2$ holds for samples of varying thicknesses, the extracted slope systematically decreases as the sample thickness increases. This trend is consistent with a surface-driven mechanism [7, 11, 25]. In thicker samples, $\sigma_{xx}$ becomes increasingly dominated by bulk channels, which exhibit negligible NLHE compared to the surface channels. Consequently, the increase in ${\sigma_{xx}}^2$ is not accompanied by a proportional enhancement in the nonlinear response, resulting in a reduced slope.

Table 1. Comparison of surface-driven NLHE parameters for various materials.

| **Materials** | **Dimension ($L_x$ x $L_y$ x thickness)** | **Input Current $I$ (μA)** | **Input Current density (μA/μm²)** | **$V_{xy}^{2\omega}$ (μV)** | **$\chi_{xy}^{2\omega}$ (μm V⁻¹)** | **T (K)** |
|---|---|---|---|---|---|---|
| BiTeCl (This work) | 14 μm × 14 μm × 12 nm | 5 | 30 | 107 | 1.68 | 300 |
| Te [25] | 30 μm × 1.7 μm × 24 nm | 5 | 120 | 175 | 0.02 | 300 |
| Bi [18] | 40 μm × 1.3 μm × 100 nm | 60 | 460 | 0.6 | 0.02 | 300 |
| $NbIrTe_4$ [11] | 27 μm × 3 μm × 15 nm | 200 | 4444 | 22.5 | 0.018 | 300 |
| $Bi_2Te_3$ [21, 22] | 25 μm × 10 μm × 10 nm | 900 | 9000 | 250 | 0.006 | 300 |
| $TaIrTe_4$ [7] | 7 μm × 1 μm 20 nm | 600 | 30000 | 100 | 0.005 | 300 |
| $Bi_2Se_3$ [20] | 230 μm × 20 μm × 20 nm | 1500 | 3750 | 13 | 0.0004 | 200 |
| $Pb_{1-x}Sn_xTe$ [23] | / | 100 | / | 250 | / | 300 |

We compare the performance of BiTeCl with other materials exhibiting surface-driven second-order NLHE, as summarized in Table 1. Our 12-nm-thick BiTeCl device delivers a substantial output voltage of $V_{xy}^{2\omega} \approx 107$ μV driven by a very low input current of only 5 μA. Although Te flakes yield a slightly higher absolute voltage, they require a significantly higher input current density (120 μA/μm$^2$) compared to BiTeCl (30 μA/μm$^2$). Crucially, BiTeCl exhibits a giant surface-driven second-order nonlinear Hall coefficient $\chi_{xy}^{2\omega}$ of 1.68 μmV$^{-1}$ at room temperature, which is about 80 times larger than that of Te (0.02 μmV$^{-1}$) and orders of magnitude superior to other topological materials. This demonstrates that BiTeCl possesses superior efficiency in surface-driven, scattering-mediated nonlinearity, enabling massive signal generation with minimal power consumption.

The unprecedented magnitude of the nonlinear Hall response in BiTeCl is a direct consequence of its unique polar crystal structure and the resulting asymmetric surface electronic states. In conventional non-polar systems with topological or Rashba-type surface states, idealized top and bottom surfaces are structurally identical but host helical spin textures with opposite chiralities due to inversion symmetry. This symmetry dictates that the nonlinear currents generated via both skew scattering and side-jump mechanisms possess opposite signs on the two surfaces, leading to a theoretical cancellation of the net signal [12, 13, 16, 33]. In practice, however, real-world non-polar samples can exhibit a finite but weak response because the top (exposed to vacuum/air) and bottom (adjacent to the substrate) surfaces experience different dielectric environments, slightly breaking the perfect symmetry.

In stark contrast, the giant response in BiTeCl arises from an intrinsic and fundamental asymmetry guaranteed by its crystal lattice and rich surface states. The cleavage of BiTeCl occurs only between Te and Cl layers. Therefore, a BiTeCl flake inherently possesses two chemically distinct terminations: one Te-terminated and one Cl-terminated, as illustrated in Fig. 3(e). This structural certainty precludes the existence of equivalent surfaces with opposing spin textures. As revealed by ARPES measurements [36, 37] and illustrated in Fig. 3(f), robust Rashba-split surface states, comprising two surface bands and one surface resonance band, are hosted on the n-type Te-terminated surface. Crucially, these coexisting states share the same spin helicity, ensuring their contributions to the nonlinear Hall effect are additive rather than competitive.

Conversely, on the opposing p-type Cl-terminated surface, the Fermi level resides within the bulk bands [42], rendering surface states inaccessible and their contribution to the surface-driven mechanism negligible. Since the Cl-terminated surface does not support the necessary states to generate a significant nonlinear current, the condition for cancellation is fundamentally absent. Consequently, the total nonlinear response is not a result of a superposition of two equivalent surfaces but is dominated entirely by the single Te-terminated surface. This unique single-sided configuration allows the full efficiency of both skew scattering and side-jump processes. The synergistic combination of structural polarity and favorable surface states ensures that BiTeCl exhibits a giant, uncompensated nonlinear Hall effect.

**4 Conclusion**

In summary, we have successfully observed a giant, surface-driven, in-plane second-order NLHE in Rashba-type polar BiTeCl crystals, which persists robustly up to room temperature. By systematically excluding bulk contributions through symmetry analysis and thickness-dependent measurements, we established that the dominant origin of this response lies in the electronic states localized at the surface. The determined nonlinear Hall susceptibility reaches an unprecedented value of 1.68 $\mu mV^{-1}$ at 300 K, exceeding that of the best previously reported surface-dominated systems by a factor of 80. We attribute this giant enhancement to the synergistic effect arising from BiTeCl's polar crystal structure and its rich, strongly Rashba-split surface states. Scaling analysis reveals that both extrinsic skew-scattering and side-jump mechanisms contribute to the observed nonlinearity. The combination of a giant, surface-driven nonlinear response and low input current requirements makes BiTeCl a promising candidate for efficient, low-power radio-frequency rectification and wireless communication technologies, offering a distinct advantage for device miniaturization.

**Acknowledgments**

This work was supported by the National Key R&D Program of China, the Ministry of Science and Technology of China, the National Natural Science Foundation of China.

**Data availability**

The data that support the findings of this study are available from the corresponding author upon reasonable request.

**Declarations**

**Conflict of interest**

The authors declare that they have no conflict of interest.

**Author contributions**

Dong Qian designed the research. Ziheng Wang and Hui Li grew the samples. Zhihua Liu carried out the transport experiments with assistance from Yongbo Lv, Ziheng Wang, Hanru Feng, Zhiwei Zhang, Bo Zhang, Feng Liu, Guohua Wang, Shengwei Jiang, and Hao Chu. Zhihua Liu and Dong Qian wrote the manuscript. All authors read and approved the final manuscript.